# Multi-radial LBP Features as a Tool for Rapid Glomerular Detection and Assessment in Whole Slide Histopathology Images

Olivier Simon[1], Rabi Yacoub[2], Sanjay Jain[3], and Pinaki Sarder[1]

[1]*Department of Pathology and Anatomical Sciences, SUNY-Buffalo*
[2]*Division of Nephrology, Department of Medicine, SUNY-Buffalo*
[3]*Department of Pediatric Nephrology, Washington University School of Medicine*

*Address all correspondence to: Pinaki Sarder*[*]*, pinakisa@buffalo.edu, (716) 829-2265*

**Abstract** – We demonstrate a simple and effective automated method for the segmentation of glomeruli from large (~1 gigapixel) histopathological whole-slide images (WSIs) of thin renal tissue sections and biopsies, using an adaptation of the well-known local binary patterns (LBP) image feature vector to train a support vector machine (SVM) model. Our method offers high precision (>90%) and reasonable recall (>70%) for glomeruli from WSIs, is readily adaptable to glomeruli from multiple species, including mouse, rat, and human, and is robust to diverse slide staining methods. Using 5 Intel(R) Core(TM) i7-4790 CPUs with 40 GB RAM, our method typically requires ~15 sec for training and ~2 min to extract glomeruli reproducibly from a WSI. Deploying a deep convolutional neural network trained for glomerular recognition in tandem with the SVM suffices to reduce false positives to below 3%. We also apply our LBP-based descriptor to successfully detect pathologic changes in a mouse model of diabetic nephropathy. We envision potential clinical and laboratory applications for this approach in the study and diagnosis of glomerular disease, and as a means of greatly accelerating the construction of feature sets to fuel deep learning studies into tissue structure and pathology.

## 1. INTRODUCTION.

Virtual histopathology, in which histology slides are digitally scanned at high resolution and stored as whole-slide images (WSIs), has rapidly assumed a prominent role in pathology research. It allows for easy sharing and storage of tissue structure information without many of the drawbacks of working with the original glass slides, such as their fragility, bulkiness, and observer variability due to illumination differences [1-3]. At the same time, the rapid evolution of machine learning methods has made image-based computer-aided diagnosis increasingly practicable, particularly for the analysis of light microscopy WSIs, dermatoscopy images, and radiological data [4-6]. Such methods include support vector machines (SVMs), boosted decision trees, artificial neural networks and non-negative matrix factorization methods [7], all of which have the potential for condensing very subtle and high-dimensional image features into relatively simple and decisive pathologic classifications [8].

The glomerulus, a histologic structure found in the cortex of the kidney, forms the initial interface for filtration of metabolic wastes from the bloodstream. Glomeruli are roughly spherical and contain tightly packed loops of capillaries juxtaposed against a convoluted collagenous sheath (the basement membrane) which is supported by podocytes and functions as the primary filter. An intervening supportive structure, the mesangium, assists in regulating blood flow and readily takes up common histological stains such as PAS [9, 10].

Given its crucial role and structural intricacy, diseases of the glomerulus are widespread and can have devastating impact. Although many varieties are known, such as IgA nephropathy and lupus nephritis, glomerular disease is a particularly common as a co-morbidity of diabetes, mainly in the form
of diabetic nephropathy (DN). In DN, thickening of the basement membrane, expansion of the mesangium and overall loss of the podocytes and filtration boundary leads first to microproteinuria and then, in approximately 4-17% of type II diabetic cases, renal failure and end-stage renal disease (ESRD) [10, 11]. Importantly, owing to the dramatic increase in diseases of glucose metabolism such as metabolic syndrome and type 2 diabetes (anticipated to afflict 37% of the US population), DN is projected to greatly increase in prevalence [12].

---

[*] *This article has not been formally edited by the corresponding author.*

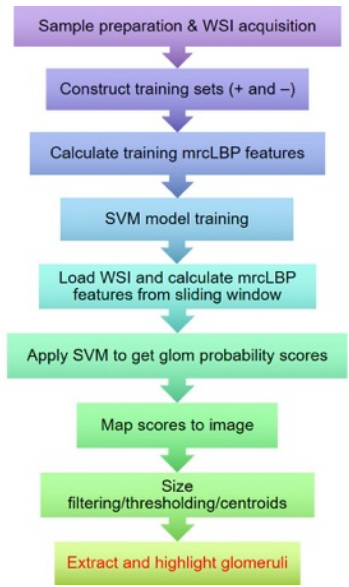

**Figure 1**. Schematic of pipeline involved in glomerular detection. Samples are first prepared, training sets are gathered by hand and mrcLBP feature vectors are calculated from them. Next, an SVM classifier is trained using these sets and deployed via a sliding window on the WSI of interest to give a map of glomerulus detection scores over the slide. This map is thresholded and filtered to remove objects too large or small to be glomeruli, centroids are calculated for the remaining dark regions, and the glomeruli are highlighted or automatically cropped.

Currently, assessment of DN depends on manual annotation by pathologists, which is time-consuming and represents a significant bottleneck and expense in the treatment pipeline. Furthermore, despite general agreement on the hallmarks of glomerular damage in DN, pathologists often disagree on scoring and diagnosis of the condition [10]. Finally, the progression of DN is subject to dramatic differences in progression rates, with some cases rapidly progressing to ESRD while others spontaneously regress [9, 11]. These factors combine to create unusual prognostic difficulties which, along with DN's large and growing prevalence, underscores the desirability of automated tools to facilitate the assessment of DN from histopathology data. Yet, though machine learning techniques have been applied to other diabetic complications, such as diabetic retinopathy [13] and diabetic neuropathy [14], and semi-automated segmentation of glomeruli [15, 16], fully supervised learning techniques for efficient glomerular segmentation and classification from WSIs remain an open challenge.

To this end, a wide assortment of textural classifiers have been mustered in the search for optimal supervised classification of image features, including histogram of oriented gradients (HOG), local binary patterns (LBP), concurrence matrices, grayscale histograms, chromaticity moments and Gabor wavelets [17, 18]. Approaches to glomerular detection have also made use of genetic algorithms [16, 19, 20] and large Internet databases such as Cytomine [21], yet these are too computationally intensive or offer limited performance. Afield of machine learning, staining with antibodies to the intermediate filament nestin can highlight glomeruli [22], while in MRI, paramagnetic nanoparticles can outline the glomerular basement membrane [23]; however, such label-based methods are laborious and suffer from high variability and noise [24].

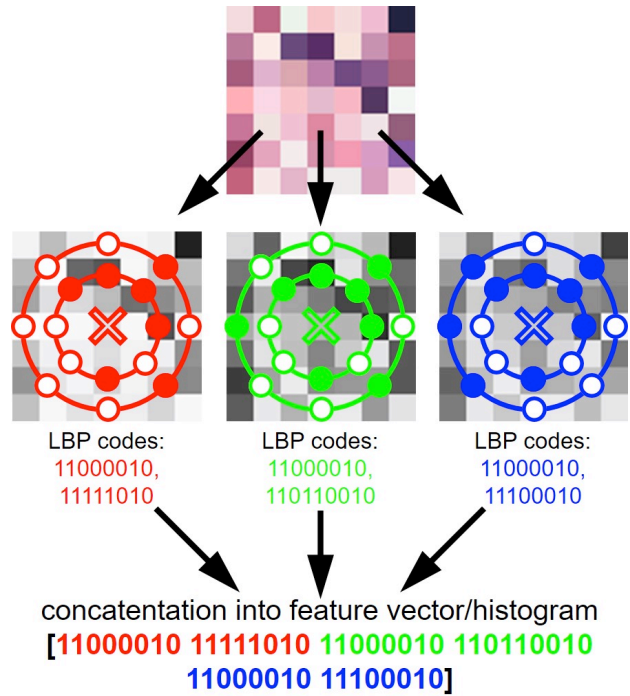

**Figure 2**. Overview of the procedure for mrcLBP feature vector construction. The LBP feature is derived from the intensity of a circular neighborhood of eight pixels thresholded to a central pixel; pixels less intense than the central pixel are assigned a value of 0, and 1 otherwise. In mrcLBP, this process is simply repeated with circular neighborhoods of multiple radii, and for each of the RGB color channels. The resulting binary patterns are then concatenated or can be combined into a histogram for training a classifier.

Here, we employ the LBP texture descriptor, first described in 1994 [25], in a supervised-learning pipeline to achieve glomerular detection (Fig. 1) and also to distinguish between disease and control glomeruli in a mouse model of DN. In brief, LBP works by defining a circular set of "neighbor" pixels at a fixed radius from each image pixel, then thresholding each neighbor with respect to the central pixel; those with intensity greater than or equal to the center pixel are thus assigned "1" values, while those with lower intensity are assigned "0"s. The resulting binary string is a simple and rapidly calculable local descriptor, which can then be binned and normalized based on the frequency of each pattern type. Because of its localized thresholding, LBP is notably robust to variations in image illumination [26]. Moreover, they can readily be adapted for "multi-resolution analysis" or "multidimensional LBP" by performing LBP at multiple radii or resolutions for each image pixel [25, 27].

In this work we use a similar approach, referred to as multi-radial color LBP (mrcLBP). Briefly, after gathering a training set of glomerular images, we calculate LBP descriptors for four different radii per center pixel, as well as for each RGB color channel, then concatenate the resulting 12 descriptors to create a final feature vector (Fig. 2). Next, these mrcLBP features are used to train an SVM classifier, which can be deployed on other WSIs for glomerular detection. If desired, a trained deep neural network model can be applied, increasing the overall precision further. Finally, by applying mrcLBP to training sets gathered from a mouse model of DN, we find evidence that mrcLBPs may be useful not only in automated detection of glomeruli, but in assessing their pathologic condition.

## 2. METHODS.

*I. Tissue Section Preparation.*

a) *Mouse slides*: Kidney tissues were gathered from 10 control and 7 STZ-treated (diabetic/DN model) mice, euthanized under institutionally approved laboratory protocols. Tissues were fixed with formalin, embedded in paraffin blocks, and cut to 5 μm slice thickness along the sagittal plane using an Olympus CUT 4060 microtome. Slices were then stained with H&E.

b) *Rat slides*: Rat tissue sections (a gift from Dr. Tracey Ignatowski, Pathology and Anatomical Sciences, University at Buffalo) were prepared using the same protocols as for mouse except that in addition to H&E, PAS, Jones silver, Gömöri's trichrome, and Congo Red (CR) staining were also carried out. All slices were cut to 5 μm thickness, with the exceptions of the Jones silver and CR slides which were cut to 2- and 8-μm thickness, respectively. These thicknesses were chosen to mimic standard clinical slide preparation practices.

c) *Human slides*: Human kidney tissue section and needle-biopsy WSIs were generously made available by Dr. Agnes Fogo, Department of Pathology, Microbiology and Immunology, Vanderbilt University Medical Center and Dr. Sanjay Jain, Division of Nephrology, Washington University School of Medicine. All slides were prepared using standard PAS staining protocols

*II. Image collection and processing.* Imaging for mice was conducted using a whole-slide bright-field microscope (Aperio, Leica, Buffalo Grove, Illinois), with a 40× objective and NA = 0.75; resolution was 0.25 μm/pixel for all acquired images. All images were initially acquired by the microscope as semi-proprietary .SVS or .SCN formats; for ease and generality of use, these files were converted to .JPG and .TIF formats, either using Aperio ImageScope to individually convert smaller groups of images, or for larger batches, the Bio-Formats command-line tool "bfconvert" (freely available from the Open Microscopy Environment, http://downloads.openmicroscopy.org/). To make the images more tractable in terms of RAM and disk usage, all WSI from human and rat were resized to 50% of original.

*III. Training set construction.* To train a classifier for glomerular detection, for each of the three species two sets of training images were collected, containing glomeruli (glom+) or no glomeruli (glom-). SVM training typically requires several hundred to a few thousand images of each class to achieve maximum classification performance [8, 28, 29]. As an excess of negative training images appears to improve precision, all training sets used contained at least twice as many glom(-) images as glom+. All training images were extracted by hand in Matlab at a size of 576 x 576 pixels. This size was chosen because it easily accommodates entire glomeruli with ample textural detail, and because it is highly composite ($576 = 2^6 \cdot 3^2$), allowing studies at a range of scales without having to round off pixels. All clearly distinguishable glomeruli were collected from each WSI, unless otherwise specified, and saved as .png files.

In total seven distinct training sets were created, comprising one glom(-) and glom(+) set each from: H&E-stained control mice; rat sections in five different staining protocols; and human biopsies and kidney sections from Washington University. In addition, one glom(+) set was collected from STZ-treated mice. Briefly, the H&E training set for mouse comprises 1059 glom(+) and 1799 glom(-) images gathered from 15 WSIs from healthy mouse kidney. The rat training set consists of glom(+) and glom(-) images gathered from from congo, H&E, Jones, PAS, and Gömorri trichrome, gathered from three different kidney slice WSI per stain, giving a total 7099 glom(+) and 15750 glom(-) images from 15 whole slides. 10 whole slides left as test images. To reduce training time, classification was found to be satisfactory using every third image out of this "universal" training set. The human training set consists of 1649 images—515 glom(+), 1144 glom(-)—gathered from a total of 25 needle-biopsy slides and 1 large tissue section from Washington University data. For all training sets, careful attention was paid to ensure that the WSIs that provided the training images were kept distinct from the testing or "holdout" images used to assess the performance of the classifier.

*IV. Multi-radial color LBP (mrcLBP).* In order to extract features from the training set image,

LBP features were applied using Matlab's "extractLBPFeatures" function with default settings (L2 normalization, one bin per image, 8 neighbors), except that the 'Upright' flag was set to 'false' and the radius is specified (default is 1 pixel). The 'upright' flag removes the rotation-invariance of the LBP feature, but also reduces its dimensionality from 59 to 10 components to afford a simpler overall feature for training and classification with little or no change in speed or detection ability. We found no difference in performance using this version with fewer components.

The mrcLBP feature vector for a given image region is constructed with the settings above, but with the following additional simple modifications: first, instead of using only the total intensity information, LBP is carried out on the separate R, G, and B channels. Secondly, instead of extracting the LBP pattern at a single radius for each channel, patterns are gathered at four different radii. Best results, particularly with regard to minimizing inter-laboratory variability between training and test images, were generally achieved using radii of 1, 3, 9, and 27 pixels, though other combinations, such as [2 4 6 8], are also viable for glomerular detection. To construct the final mrcLBP feature vector, the extracted LBP vectors for each combination of radius and color channel are concatenated (R1 R3 R9 R27 G1 G3 G9 G27 B1 B3 B9 B27). The result is a 120-dimensional feature vector containing information on textural variations in terms of local intensity differences as a function of color and at over a range of scales. All experiments involving feature extraction, classifier training, and feature detection were carried out using a pool of 5 Intel(R) Xeon(R) CPU E5-2697 v3 cores @ 2.60GHz, with 40GB RAM.

*V. Feature extraction and SVM training.* mrcLBP feature vectors were extracted from a cropped central region (typically 1/6 the overall width) of the chosen training images using a parallelized loop and concatenated them to form an array, together with a class label vector indicating which of the feature vectors in the array correspond to which classes. Groups of positive and negative training images examples are processed in that order, so that for a training set of m positive and n negative training images, the label array consists simply of m 1's followed by n 0's. Once both feature and label vectors have been prepared from the training set, SVM training is carried out in the usual fashion, in this case using Matlab's fitcsvm() function, with a linear kernel and 'boxConstraint'=1; polynomial kernels and other non-default arguments were not found to yield substantial improvements in classification performance.

*VI. Sliding-window extraction of mrcLBP features from test image.* For glomerular detection, the WSI image of interest was fed to the trained SVM classifier by way of a standard sliding-window approach (row-by-row, left-to-right). A list of locations on a square grid to be visited by the sliding window was generated with an adjustable stride (typically 64 pixels), and corresponding regions were successively cropped and fed to the SVM using a parallelized loop. Window size is identical to the size of the training images (576x576), as is the cropping step. In this phase, two main adjustable parameters are involved: the step size, which determines the separation between successive window positions, and a resolution calibration factor that ensures that the resolution ($\mu$m/pixel) of the glomeruli in the training images is commensurate with that of the glomeruli in the WSI being scanned. In cases where the glomerular training set was collected from WSIs from the same instrumentation, or where glomerular pixel size is closely similar to that of the training images, the calibration factor can be left at 1; for the step size, we found that a stride of 45 pixels provided a good trade-off between detail and speed, though larger values can be used for successful detection.

*VII. Application of trained SVM to WSI.* The trained SVM classifier was deployed on the concatenated mrcLBP feature vectors from each of the sliding window locations, resulting in a map of scores indicating the classifier-estimated likelihood that the corresponding window position belongs to the glom(+) or glom(-) class. This score map, in which darker pixels correspond to a higher probability that a glomerulus is present, can then easily be thresholded and overlaid on the original WSI.

*VIII. Score map processing.* To remove large regions of similar texture to the glomeruli (particularly the kidney medulla and certain unusual growths) which can cause false positives, the raw score map is first thresholded by intensity, then size-filtered using Matlab's bwareaopen() function to remove any dark objects in the map that are too large or small to be glomeruli, and thresholded a second time to better isolate glomeruli. A final constant application of bwareaopen() is then useful to remove stray single pixels. In most cases, the first threshold was left at 0, while the glomerular area range is dependent on the resolution of the WSI and the stride of the sliding window positions visited but is typically 6-50 pixels. The final threshold typically takes a value between 0.15 and 0.35 and once set by trial and error can be applied to many slides without modification. Finally, to minimize redundant positives, Matlab's regionprops() tool is used to locate the centroids of the remaining regions. These are used to display and automatically crop the glomeruli from the WSI as 576x576 images, which are saved to a folder.

*IX. Deep learning model training for result refinement.* For precision improvement of multi-stain glomerular detection, a GoogLeNet classification model was trained in DIGITS with Nesterov's Accelerated Gradient, using the full rat training set of 7099(+) images, 15750(–) images. Of these, 20% were randomly assigned to a validation set, and 10% to a test set. Applying this model to the test set, the false positive rate was 3/1575.0 = 0.0019%, and the false negative rate was 3/709.9 = 0.0042%. Both training and deployment of the GoogLeNet model was carried out using an Nvidia GeForce 1080 GPU.

*X. Performance testing of glomerular detection.* In order to reasonably judge the generality and robustness of the detector, glomerular images being tested are kept distinct from those on which the classifier was trained. Therefore, for each type of experiment using a series of WSIs, specific entire WSIs were denoted in advance as either sources for training set construction, or as "holdouts" for subsequent testing of glomerular detection. These holdout images, as well as the WSIs used for construction of the training sets are available in the documentation in the file "PositivesNegativesHoldouts.xlsx". After glomeruli were hand-counted to establish ground truth and the SVM false positives were counted, false negative, recall, precision and F1 scores were calculated directly.

*XI. Characterization of STZ versus control glomeruli.* Differences in mrcLBP scores between STZ and non-STZ mice were studied using a method closely similar to that employed in glomerular detection, except that instead of training the SVM on images with or without glomeruli present, the training image sets consisted of glomeruli collected by hand from either control or STZ-treated mice. To assess the robustness of this approach, SVM classifiers were trained using either 10 random cross-validation runs (100 holdout images apiece), and the mean score values of the test images in each case were calculated. The resulting SVM scores were aggregated and displayed as a histogram for both the STZ- and non-STZ holdout images.

## 3. RESULTS.

a. *Glomerular detection in mouse.* Examples of the mrcLBP feature pipeline applied to mouse WSIs are shown in Figure 3a-c. Although there was some variability in the performance of the method between slides, all showed precision for glomerulus detection in the range of 0.83-0.98, with an average of just over 90% (in terms of total glomeruli in all slides). Recall and F1 scores were somewhat lower, at 80% and 85%, respectively. Visual inspection of the SVM score maps confirms that the vast majority of positive locations contain glomeruli (Fig. 3a, green boxes). Connected areas of the WSI with strong textural similarity to the glomerulus, specifically parts of the kidney medulla, are removed by a simple size-filtering and thresholding approach, ensuring that false positives do not overwhelm the true positives (Fig. 3b).

Notably, this detection pipeline is robust even to the presence of certain pathological changes within the tissue, as shown by the reasonably good precision achieved even when the pipeline is trained on glomeruli from control mice and applied to glomeruli from STZ-treated mice—85% precision for STZ (n=3 WSIs), versus 93% precision when the same trained classifier is applied glomeruli from control mice (n=7 WSIs), with standard deviations of 0.0188 and 0.072 respectively (Fig. 3c). Observed recall for glomerulus detection, in fact, showed no significant difference between control and STZ mice, with values of 78% and 82% respectively and standard deviations of 0.106 and 0.0461.

b. *mrcLBP performance in rat, with multiple stains.* Representative results of mrcLBP-based glomerular detection in rat kidney are presented in Figure 4a, for a classifier trained on an equal mixture of glom-positive and glom-negative images from five separate stains. As can be seen in the zoomed regions, the resulting pipeline is robust to variations in stain type, with abundant glomeruli detected in all five cases. Figure 4b shows a summary of results for precision, recall, and F1 score for each of the five types of stained test images (2 WSI each). Total ground truth values for each stain type are 958 glomeruli for congo; 989 for jones; 927 for H&E; 881 for PAS,
and 909 for Gömorri trichrome, yielding a total of 4664 glomeruli, with 3450 total positives.

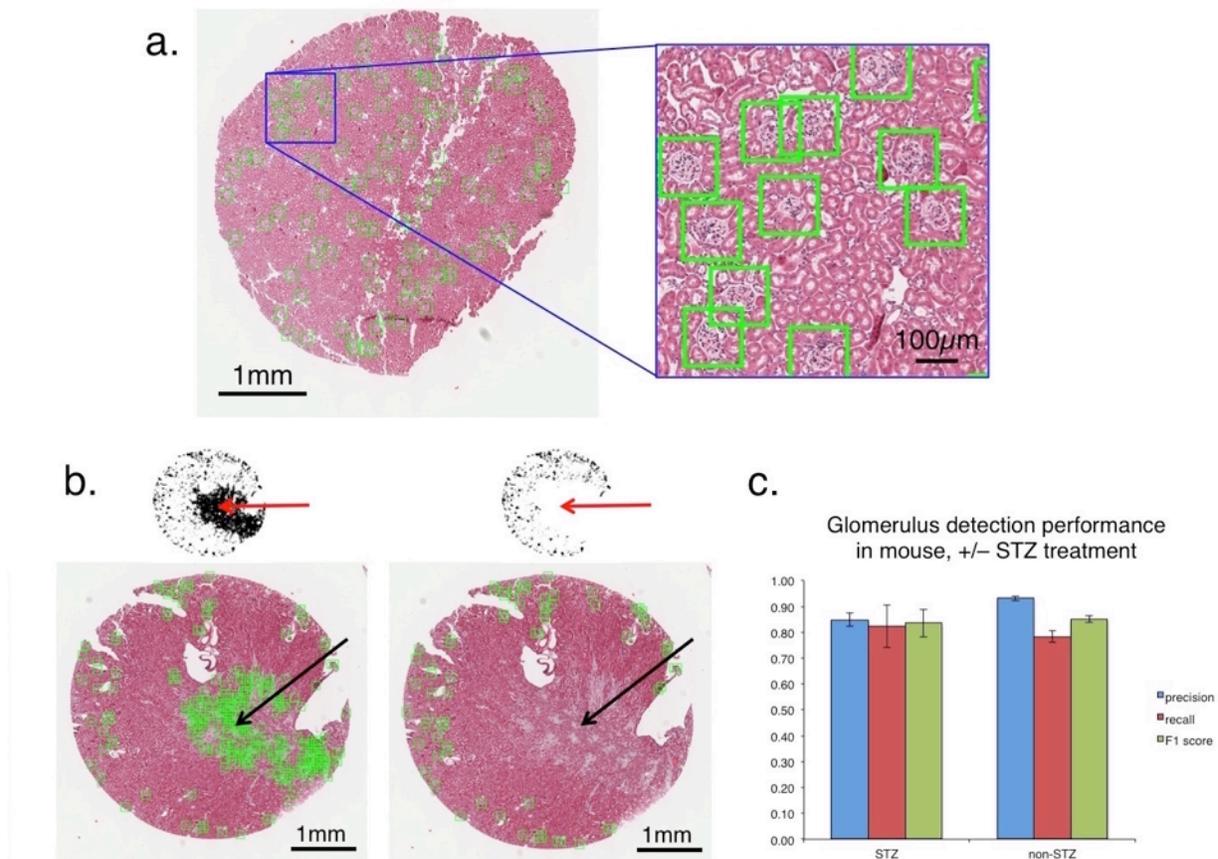

**Figure 3**. Performance of the mrcLBP glomerular detection pipeline in mouse WSIs. **a)** Representative mouse WSI (P25-2) with glom locations automatically marked (green squares), H&E stain. **b)** Example of removal of large connected areas from the SVM output using area opening. Left, mouse WSI prior to area opening, showing scores map (top) and prospective glomeruli in green boxes (bottom). Right, same slide after successful opening, showing removal of false positives due to medullar texture at the center-lower-right of the sample. H&E stain. **c)** mrcLBP-based glomerulus detection in mouse WSI, with varying degrees of STZ treatment, for a classifier trained on non-STZ glomeruli. Although STZ causes noticeable histological changes in certain glomeruli, the performance of the detection algorithm is only minimally affected, with similar F1-score for all three cases.

Congo stain showed poorer precision than the other types, possibly due to the much lower visual contrast of the glomeruli in these images (detection is difficult even by eye), though the recall was similar. Jones and H&E slides both showed the highest precision, above 90%, while trichrome-stained slides presented slightly lower precision (88.7%) but offered the highest recall, with 73% of the ground truth gloms detected. Much as the lower precision values for Congo stain are likely due to the very low visual contrast it provides in the glomeruli, the higher recall with trichrome stain is likely a function of the very striking color differentiation it achieves, with glomeruli typically greenish against the red background of tubules and blood vessels.

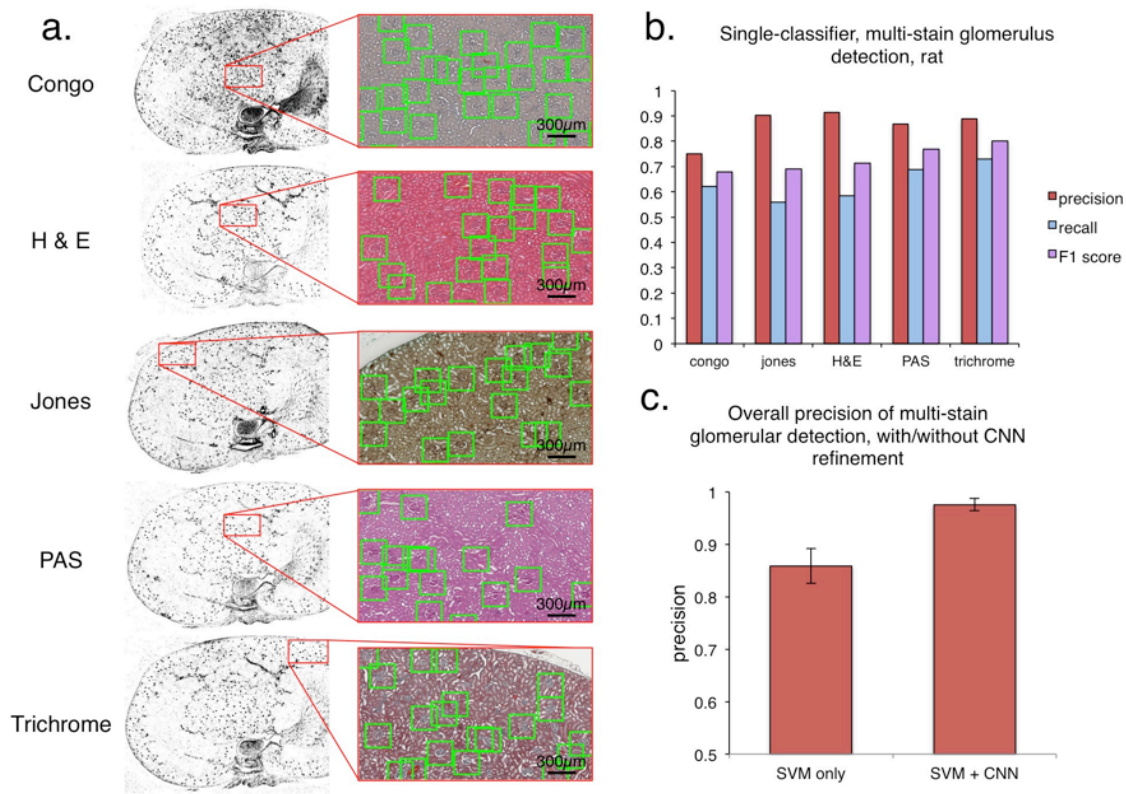

**Figure 4**. Performance of the glomerular detection pipeline for rat WSIs, with five stains. **a)** Images showing glom detection results for 5 different stains. From top to bottom: congo, jones, H&E, PAS, and Gömorri trichrome. In this case, an SVM classifier was trained simultaneously on equal numbers of examples from all five stains. Glomeruli are visible as black spots in the score maps (left column) and with overlaid boxes over the original color image (right column). **b-c)** Performance of the mrcLBP method when trained on multiple stain types simultaneously, with or without purification by a deep neural network. Top (b), performance of the method after SVM classifier alone in 10 rat WSI. To increase recall, score maps were thresholded lower, admitting more total positives. Bottom (c), result of purifying the collected glomerular images with a GoogLeNet neural network model trained on 7099 rat glomeruli in five stains. Precision now exceeds 97.5% for all stains, with minimal decrease in recall (not shown).

c. *Precision improvement of harvested rat glomeruli using neural network classifier*. Although neural network approaches can be extremely versatile and achieve very high precision in certain tasks, our experience showed that CNN's (at least readily available multi-purpose ones such as AlexNet or GoogLeNet, without any special architectural modification for glomerular detection) could achieve classification accuracies of up to 99.8% when trained on thousands of rat or mouse glomeruli. Though impressive, this performance level is still not adequate for glomerular detection by sliding-window analysis of an entire WSI, which may visit tens of thousands of window positions and could therefore still lead to hundreds of false positives. Conversely, many applications would require a precision of much greater than the 90-93% precision values we obtained using mrcLBP with SVM alone. We therefore combined the approaches in tandem, using the mrcLBP feature and SVM first to gather a high-quality set of candidate glomeruli, followed by a trained GoogLeNet model to "purify" the set further.

The results of this tandem purification are summed up in figure 4c. Notably, while the recall and F1 values were negligibly decreased by adding the additional classivier, the overall precision reached 97.5% (2929 true positives, versus only 74 false positives), or a 6-fold decrease in the proportion of false positives. The lowest-precision stain was now Jones (95.3%), surpassed by congo (96.6%), trichrome (98.2%), H&E (98.4%) and, highest of all, PAS (99.1% precision). Running the neural network classifier with an Nvidia GeForce 1080 GPU, purification of the glomeruli from each stain was achievable in under a minute.

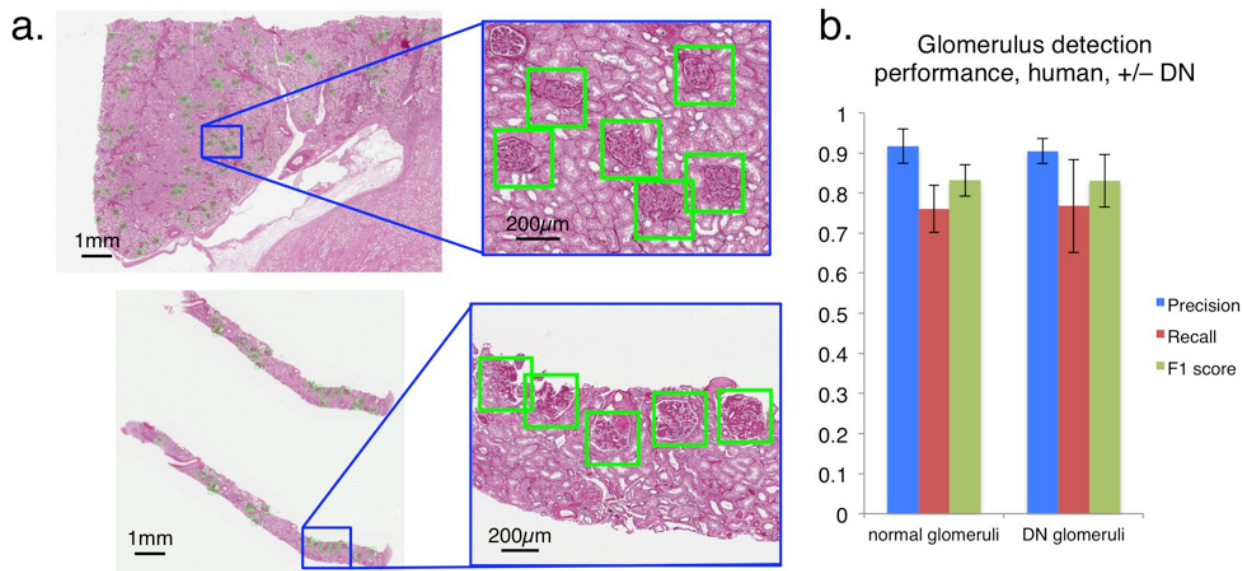

**Figure 5**. Examples and results from glomerular detection in human WSIs using mrcLBP. **a)** Examples of successful application of mrcLBP to human renal WSIs, with detected glomeruli (green boxes). Top, renal section. Bottom, needle biopsy. PAS stain. **b)** Results of detection of glomeruli in human tissue samples. Left group, non-DN patients, from 9 slides containing 1088 glomeruli. Right group, result from DN patients, from 5 slides containing 99 glomeruli.

    d. *mrcLBP: application to human glomeruli*. Applicability of a detection method to human disease data is among the most crucial criteria for clinical and research significance. Therefore, having confirmed the viability of mrcLBP for glom detection in mouse, rat and a variety of staining protocols, we sought to examine its performance in needle biopsies and tissue sections from human kidney. Our training data comprised samples from 1 healthy individual, 7 individuals with diabetic nephropathy, and 1 with IgA nephropathy. Examples of glomerulus detection using the resulting SVM model to test WSIs are shown in figure 5a, for a needle biopsy (DN) and tissue section (normal kidney).

    Figure 5b shows basic statistics for glomerular detection, on slides from 5 patients with DN (n = 5 WSIs) and 3 patients with normal glomeruli (n = 9 WSIs). As may be expected for a classifier trained on a mixture of DN and normal glomeruli, the results between the two groups are statistically indistinguishable. Specifically, overall precision values are 91.7% and 90.4% for normal and DN, respectively, with standard deviations (by slide) of 0.083 and 0.061; recall values were 76.1% and 76.7%, standard deviations 0.12 and 0.23; and F1 scores, 0.832 and 0.831, standard deviations 0.0787 and 0.131. Therefore, overall performance of the mrcLBP-trained SVM approach in human glomerular detection appears broadly similar to that demonstrated in mouse and rat WSI.

    e. *Population differences between glomeruli from control and STZ-treated mice*. We observed that the mrcLBP-trained SVM classifier, even when trained to detect both DN and healthy glomeruli, still does not detect fully sclerosed ("obsolete") glomeruli in human tissue. Therefore, we wondered whether this selectivity could be of use in classifying glomeruli by disease state. Due to the readier availability of glomerular images from mouse, and the highly controlled conditions attainable with the mouse STZ model of DN, we decided to re-examine the mouse WSI data using an mrcLBP-trained SVM pipeline, but this time with the aim of classifying harvested glomeruli according to their disease state rather than of glomerular detection.

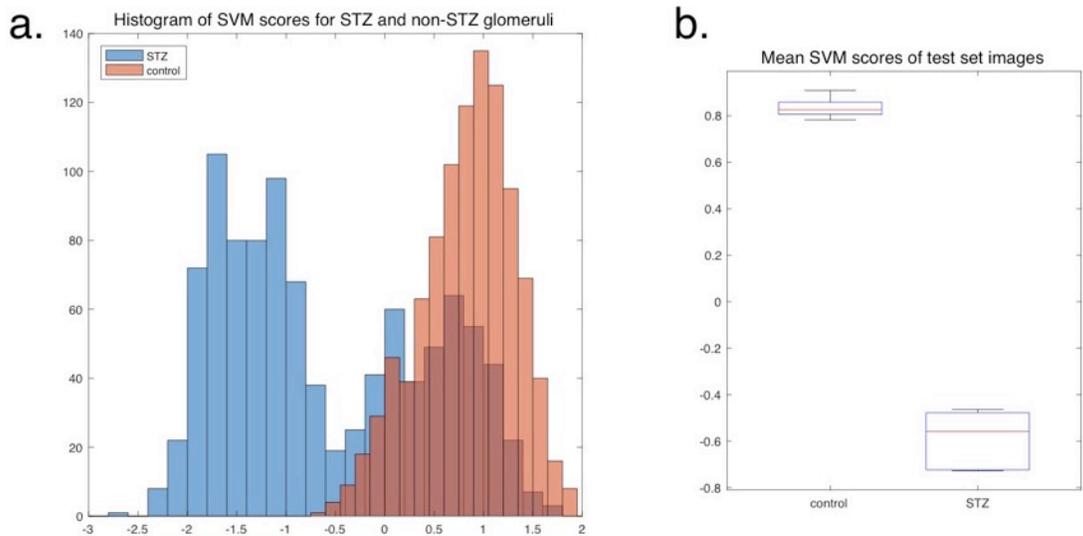

**Figure 6**. Application of the mrcLBP to a mouse model of diabetic nephropathy reveals significant textural alterations. **a)** Distribution of aggregated classifier scores after ten random cross-validations. Holdout images from STZ-treated mice (blue bars) show a strongly bimodal distribution of SVM scores, while holdouts from non-STZ-treated mice (orange bars) are unimodal, consistently receiving positive classifier scores. **b)** Boxplot showing spread of the mean values of STZ and non-STZ scores from all cross-validations.

Using randomized cross-validations of the mouse training data, we examined the distribution of scores produced by an SVM trained to distinguish between glomeruli from control and STZ-treated mice. The aggregated results from 10 such cross-validations (100 holdout images from each class per cross-validation) are presented in figure 6a-b. A striking difference in the distribution of the SVM scores for control and STZ-treated mice is immediately apparent in the histogram in figure 6a; control glomeruli are almost universally assigned high SVM scores, creating a single large peak to the right of the histogram, while STZ glomeruli are mainly located in a peak at the left of the histogram, corresponding to highly negative SVM scores. Interestingly however, there is also pronounced bimodality in the STZ distribution, with a substantial minority of STZ glomeruli found in an additional peak that closely overlaps that of the control glomeruli. This strongly suggests that the glomeruli in STZ mice belong to at least two distinct sub-populations: one showing pathologic changes and the other remaining (texturally) normal.

Even given the significant bimodality of the STZ scores, the mean values of the STZ and control score distributions show consistent and highly significant separation, as visualized in figure 6b. We therefore propose that mrcLBP-trained SVM classifiers are capable of distinguishing, given a sufficient sample size, between control glomeruli and those from STZ mice.

## 4. DISCUSSION.

In the foregoing, we have outlined and demonstrated a simple pipeline for the rapid and fully automated extraction of glomeruli from WSI. Our method is easy to implement, requires only a single classification step to achieve performance comparable or superior to the current state-of-the-art, and can process large numbers of WSIs in a short period. Due to its simplicity, a relatively small number of parameters are needed: two intensity thresholds for extraction of renal medulla and for glomeruli, and a size range for filtration of large connected components. We have also shown that the method can be combined with other classifiers, particularly deep neural networks, in order to produce very high precision glomerular detection. Moreover, we have been able to arrive at preliminary results strongly indicating a statistical signature for diabetic nephropathy in a mouse model of the disease.

In contrast, relatively few works so far have directly addressed the problem of glomerular detection and classification simultaneously. Most reported approaches have shown relatively low precision, have required access to extremely large datasets [20, 21], or have made use of biochemical markers. Regarding the latter, in addition to the immunohistochemical and MRI methods earlier alluded to, one particularly novel example involves the use of Fourier-transform IR spectroscopy to assess the progression of diabetic nephropathy in pre-selected glomeruli [30]; however, the sample size was relatively small (68 glomeruli), and the method is likely not conveniently applied to glomerular detection as such.

Other approaches have typically focused on unsupervised morphological processing strategies, for instance edge patching and contour extraction by genetic algorithm [16, 19]. In [31], glomerular diameter measurement was achieved, using a relatively complex unsupervised approach consisting of a combination of median filters, particle size filters, convex hull operations, and aspect ratios. However, only a small number of glomeruli were tested and the process is slow, requiring ~2 minutes to detect and draw a single glomerular boundary. Most such approaches are furthermore dependent on the visual distinctness of the Bowman's space, something which is readily lost in thicker tissue sections.

Possibly most notable among the methods developed to date is the "segmental histogram of gradients" (s-HOG) descriptor in *Kato et al*. [15], a modification of the standard rectangular histogram of gradients (r-HOG) feature vector. The principal disadvantage for r-HOG in glomerular detection is its very high rate of false positives; in contrast, s-HOG is claimed to correct these limitations by using HOG binning regions that are not rectangular but instead individually tailored to glomeruli. The process of constructing the s-HOG feature is however relatively complex, requiring 3 successive SVM steps. In particular, the process requires that glomeruli be first segmented before the s-HOG descriptor can be properly constructed, requiring a radial sliding-window and classifier step to estimate the glomerular boundary at multiple orientations.

Further, although the s-HOG method—in combination with a more efficient segmentation algorithm—achieves improved performance over r-HOG, particularly in terms of false positive rates, it is not an actual replacement for r-HOG, as the latter is still necessary for the preliminary SVM stage. Two additional SVMs are required to segment the glomeruli and then eliminate the excessive false positives generated in the first step. This complexity and repetition appears likely to be computationally intensive, although speed estimates are not provided by Kato et al [15]. In contrast, our LBP-based method produces few false positives to begin with, and succeeds using a feature vector of much lower dimensionality (120 vs. 512 for their r-HOG) and only a single SVM step.

By way of performance comparison, the claimed precision, recall, and F-measure values of the s-HOG method were 87.4%, 89.7%, and 0.866. This precision therefore falls slightly short of our method's typical range of 90-93%, and while our method's F1 scores and recall values fall slightly below those reported for s-HOG, we believe that our method's much greater simplicity, as well as the special importance of detector precision in applications such as the construction of training sets and the gathering of glomeruli for clinical assessment, render it a potentially superior alternative.

More recently, Laurinavicius et al. have announced success with glomerular detection using out-of-the-box convolutional neural network architectures [32]; yet their training process requires up to 33 minutes and the use of a GPU, and they do not show glomerular detection results from WSIs, but rather successful classification of previously cropped glom(+) and glom(-) images. As we have noted, while it is straightforward to develop high classification performance using such neural networks, the large number of sliding window positions that must be visited to cover an entire WSI compared to the number of glomeruli typically present means that classifiers that are excellent for sorting groups of pre-cropped images will report many false positives if deployed on WSIs directly.

Our approach is perhaps notable in that it relegates more complex machine-learning processes such as neural networks to the role of refinement as a way of "purifying" the results obtained by simpler methods. Indeed, in many cases the number of false positives is small enough that for some purposes they may be disregarded, or removed manually with good speed. As an additional limitation of the CNN approach, we discovered that neural networks were generally of little use in discerning STZ from control glomeruli, tending to over-train, whereas the arguably far "simpler" feature vector-plus-SVM system was able to find a salient difference in the two groups that is robust to cross-validation and retraining.

The most basic distinction among the approaches to automated WSI feature extraction may be usefully described as between "structural", and "textural" approaches. Texture has long been considered as an important classification feature in medical diagnostics, but mainly with reference to radiotherapy, CT, and MRI scanning [5, 33, 34] or, when applied to histopathology, mainly as an aid to cancer grading and the detection of metastases [4, 6, 35]. Texture features have the specific advantages of being simple to extract and "when extracted locally, robust to geometric and illumination changes", as opposed to structural features which typically require a segmentation step [4]

One main disadvantage of supervised learning is the risk of over-training to a particular training set; indeed, it is "…currently unknown if more generic pattern recognition approaches could recognize tissue structures across several histology labs, which would be of great practical value" [21]. In our case, all training sets were derived from the same institution or lab as the images that were subsequently used for testing. We have achieved some initial success in applying our method across laboratories, but the difficulties of accessing sufficient good quality human kidney samples from a wide range of these prevented us from attempting to create a broader, laboratory-invariant training set analogous to our stain-invariant training set. Based on our success with the latter, however, as well as the steadily improving standards for WSI acquisition and color normalization, we believe this is a readily attainable aim. Also, although our SVM can be specifically *trained* to detect the difference between STZ and control glomeruli using mrcLBP, the fact that a classifier trained on only control glomeruli (or half of each) detected STZ glomeruli almost as well as control ones suggests that the method does not overlook pathological or otherwise unusual glomeruli; this robustness certainly calls for further testing however.

A number of future developments of this work suggest themselves. One is glomerular segmentation, since the feature score maps used to detect glomeruli in our approach often already reveal the rough shape of the individual glomeruli. However, increasing the resolution to the point of useful segmentation of the glomerular boundary requires reducing the size of the image patch being classified to the point where it becomes statistically unreliable, therefore it is likely that to proceed further in this direction, additional tools will be needed to refine the very rough glomerular outline currently afforded in our score maps. Other variations of the LBP feature may lead to more efficient use of the color information, such as opponent-color LBP (OCLBP), Robust LBP (RLPBP) [26].

Finally, it is worth noting that in the case of the glomerulus, the strongly distinctive appearance of the glomerular texture relative to surrounding tissue offers an unusual advantage; many other segmentation and detection problems in medical imaging and diagnostics will not allow for this simplifying premise. Nevertheless, by demonstrating a path to glomerular detection and classification that can be run even on a home computer, or can aid in extracting large training sets for more intensive statistical analyses, we envision that our results may help contribute to new applications of computer-aided-diagnoses in pathology.